\documentclass[superscriptaddress, nofootinbib, preprint]{revtex4}[12pt]

\usepackage[utf8]{inputenc}
\usepackage{graphicx}
\usepackage {amsmath, amssymb, rotating}
\usepackage{subcaption}
\usepackage{braket}
\usepackage{footnote}
\usepackage[usenames,dvipsnames]{color}
\usepackage{mdwlist} 
\usepackage{slashed}
\usepackage{verbatim}
\usepackage{mathrsfs}
\usepackage{caption}
\usepackage{ragged2e}
\usepackage{hyperref}

\begin{document}

\title{Firewalls in General Relativity}

\author{David E. Kaplan $^*$}
\affiliation{Department of Physics \& Astronomy, The Johns Hopkins University, Baltimore, MD  21218}

\author{Surjeet Rajendran $^*$}
\affiliation{Berkeley Center for Theoretical Physics, Department of Physics, University of California, Berkeley, CA 94720}
\email{ALICEandBOB@jhu.edu}

\date{\today}
\begin{abstract}
We present spherically symmetric solutions to Einstein's equations which are equivalent to canonical Schwarzschild and Reissner-Nordstrom black holes on the exterior, but with singular (Planck-density) shells at their respective event and inner horizons. The locally measured mass of the shell and the singularity are much larger than the asymptotic ADM mass. The area of the shell is equal to that of the corresponding canonical black hole, but the physical distance from the shell to the singularity is a Planck length, suggesting a natural explanation for the scaling of the black hole entropy with area. The existence of such singular shells enables solutions to the black hole information problem of Schwarzschild black holes and the cauchy horizon problem of Reissner-Nordstrom black holes. While we cannot rigorously address the formation of these solutions, we suggest plausibility arguments for how `normal' black hole solutions may evolve into such states. We also comment on the possibility of negative mass Schwarzschild solutions that could be constructed using our methods.  Requirements for the non-existence of negative-mass solutions may put restrictions on the types of singularities allowed in an ultraviolet theory of gravity.
\end{abstract}

\maketitle

\section{Introduction}
\label{sec:intro}
The black hole information problem seemingly causes a conflict between general relativity and quantum mechanics at energy scales where both theories are known to be valid \cite{Hawking:1974sw}. Efforts to explain this mystery by relying on the physics of the emitted Hawking radiation have been shown to be inconsistent \cite{Mathur:2009hf, Almheiri:2012rt}, raising the specter of a firewall,  a singular surface at the location of the black hole’s horizon. A singular firewall would not only resolve the information problem by causing the breakdown of general relativity at the scale of the horizon, but could also potentially house the degrees of freedom necessary to explain the large entropy of the black hole and its scaling with its area. Naively, such a firewall does not seem possible - the singular energy density on the firewall would suggest a total black hole mass that is parametrically larger than the mass inferred from the Schwarzschild radius of the black hole. 

There is however a need for a firewall in a closely related scenario in classical general relativity, namely, the Reissner-Nordstrom and Kerr geometries. Both these geometries possess a Schwarzschild-like outer horizon but also contain an inner horizon. The inner horizon also exists in a region of low curvature, where classical general relativity should hold. However, this inner horizon is a source of trouble. Time-like curves from the exterior space-time can fall into the region within the inner horizon and remain in parts of space-time with low curvature. But, these curves cannot extend back into the original space-time without violating causality. Mathematically, they can only be extended into a different universe - a possibility that is physically dubious.

There has been considerable effort to eliminate this problem, centered around the idea that the inner horizon is a region of instability. For example, external perturbations that fall into the black hole are arbitrarily blue-shifted as they approach the inner horizon, a phenomenon known as {\it mass inflation} \cite{Poisson:1989zz}. It was believed that these amplified perturbations will destroy the inner horizon, leading to a singular region. 
However, recent calculations have shown that a red-shift in the perturbations caused by a positive cosmological constant would remove the divergent blue-shift, inhibiting the ability to destroy the inner horizon and rejecting this general instability \cite{Cardoso:2017soq}.  It is also clear that once one cuts off the propagation a Planck distance away from the inner horizon, the large but finite proportional blue-shift cannot guarantee the destruction of this surface\footnote{Once regulated, the energy amplification factor is $\sim M/M_{pl}$ where $M$ is the mass of the black hole. A firewall at the inner horizon would imply a local mass $\sim M^2/M_{pl}$ - thus the amount of energy in the external perturbations must be $\sim M$, the mass of the initial black hole, to create this singular region from external perturbations alone.}.  Nevertheless, this phenomenon highlights an important point: the blue-shift experienced in the interior of black hole geometries can lead to significantly higher local energy densities without affecting the exterior geometry. In colloquial terms, the increased positive energy of the local matter is balanced by  negative  binding energy without changing the net positive Arnowitt-Deser-Misner (ADM) mass observed at infinity. This raises the intriguing possibility that a singular shell, or `firewall', could be supported both at the inner horizon of a Reissner-Nordstrom/Kerr black hole and the horizon of an exterior Schwarzschild black hole, where internal negative  binding energies mask the positive energy of the shell.  In this paper, we show that this is indeed possible. 

In this paper, we construct classical firewall solutions in General Relativity by matching different spherically symmetric solutions at a shell using the Israel junction conditions.  We place the shells near the inner or outer horizon locations of the external black holes. These general solutions can be taken to the limit where the energy-momentum tensors on these shells are of Planckian density.  We first construct a classical firewall at the inner horizon of a Reissner-Nordstrom black hole -- a time-like surface whose local density is $M_{pl}^3$ and is of macroscopic size, and yet is a physical Planck length from the inner singularity (at coordinate $r=0$).  We then produce a similar construction just outside of what would be the horizon of a macroscopic Schwarzschild black hole -- here again, the interior solution admits a central singularity only a Planck distance away.  In the Schwarzschild case, the shell's energy-momentum tensor satisfies the dominant energy condition.  In both cases, we find the local mass of the shells to be much larger than the ADM mass of the exterior black hole.  We speculate on a formation story for these objects as the result of evolution from normal black holes.  Finally, we present negative-mass Schwarzschild metrics and give conditions on matter that would be required to excise these solutions. 


This yields  a calculable framework to describe classical firewalls in general relativity. The surfaces at the inner horizons of Reissner-Nordstrom geometries effectively address the geodesic incompleteness problem of their interiors without requiring the addition of an infinite number of universes.  We expect similar phenomena to hold in Kerr geometries - potentially with a counter rotating firewall near its inner horizon. The surfaces at the outer horizons of Schwarzschild (and easily extendible to Reissner-Nordstrom) geometries contain enough mass to match the perceived entropy of standard Schwarzschild black holes. In principle, standard black holes could evolve to states like these much faster than a Page time, thus suggesting a resolution to the black hole information paradox.  

\section{Reissner-Nordstrom Black Hole}
\label{sec:RN}

\begin{figure}[h!]
\centering
\includegraphics[width = \textwidth]{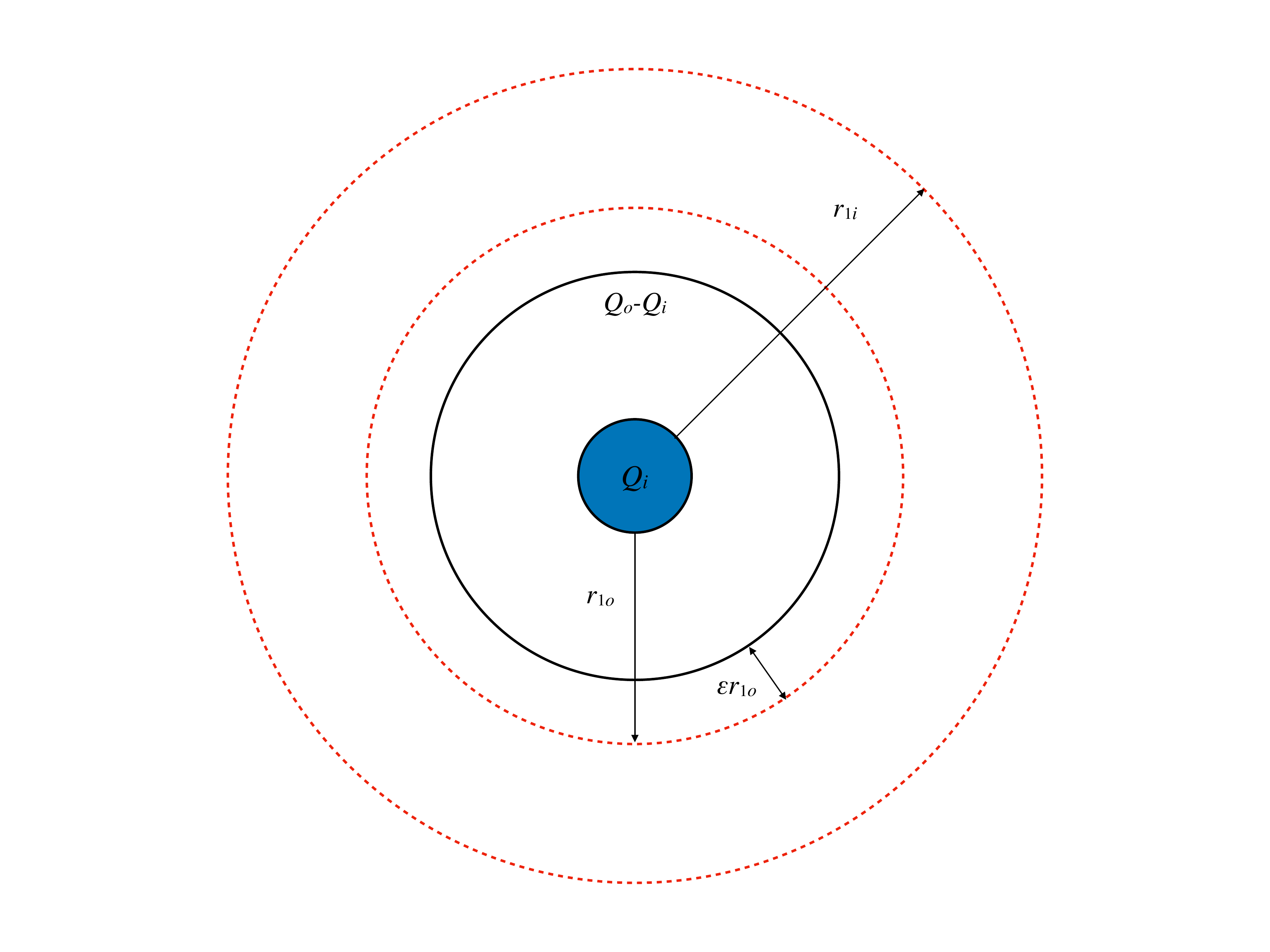}
\caption{The space-time geometry of a Reissner-Nordstrom black hole of charge $Q_o$. The interior is replaced by the space-time of a different Reissner-Nordstrom black hole of charge $Q_i \gg Q_o$. A shell of charge $Q_o - Q_i$ is placed just within the inner horizon $r_{1o}$ of the required Reissner-Nordstrom black hole (of charge $Q_o$). The inner horizon of the interior black hole $r_{1i}$ is well outside the location of this shell. The metric outside the shell is the Reissner-Nordstrom metric of the required black hole, while the interior is that of the assumed Reissner Nordstrom black hole. }
\label{fig:RNInterior}
\end{figure}


The  problems associated with the interior of a Reissner-Nordstrom black hole cannot be avoided without a singular surface close to its inner horizon.  We now show how to construct such a surface located a Planck length from the inner horizon  without changing the ADM parameters of the black hole. Our setup is depicted in Figure \ref{fig:RNInterior}. 


We assume that the interior of this black hole is described by a Reissner-Nordstrom metric whose inner and outer horizons, at $r_{1i}$ and $r_{2i}$ respectively, are much larger than the inner and outer horizons $r_{1o}$ and $r_{2o}$ of the target (exterior) black hole. We place a shell just within  $r_{1o}$. This is a time-like shell in both geometries and we can use global co-ordinates $\left(t, r, \theta, \phi\right)$ to cover the entire space-time. In these co-ordinates, the shell is at the point  $r_0 = r_{1o}\left(1 - \epsilon\right)$ just within the inner horizon of the target black hole,  with a normal vector pointing radially outward.

To match these two metrics across the shell, we take the interior metric to be: 

\begin{equation}
    g_{in} = - C^2 \Delta_{i}\left(r\right) dt^2 + \frac{dr^2}{\Delta_{i}\left(r\right)} + r^2 d\theta^2 + r^2 \sin^2{\theta} d\phi^2
\end{equation}
where $\Delta_{i}\left(r\right) = \frac{\left(r-r_{1i}\right) \left(r - r_{2i}\right)}{r^2}$ and $C$ is a constant re-definition of  local clocks in the interior metric in order to facilitate the matching of the metrics across the surface. The exterior metric $g_{out}$ is taken to be the usual Reissner-Nordstrom metric: 
\begin{equation}
    g_{out} = - \Delta_{o}\left(r\right) dt^2 + \frac{dr^2}{\Delta_{o}\left(r\right)} + r^2 d\theta^2 + r^2 \sin^2{\theta} d\phi^2
\end{equation}
where $\Delta_{o}\left(r\right) = \frac{\left(r-r_{1o}\right) \left(r - r_{2o}\right)}{r^2}$. Matching these two metrics at the point $r = r_{0}$ yields
\begin{equation}
    C = \frac{\gamma_{i}}{\gamma_{o}} \quad \mathrm{ where } \quad \gamma_i =\sqrt{ \frac{r_{0}^2}{\left(r_{1i} - r_0\right)\left(r_{2i} - r_0\right)}}, \gamma_o = \sqrt{\frac{r_{0}^2}{\left(r_{1o} - r_{0}\right) \left(r_{2o} - r_0\right)}}
\end{equation} 
and a required stress tensor $T^{\mu}\,_{\nu}$ on the surface of
\begin{equation}
\label{StressBH}
T^{\mu}\,_{\nu} = {\left(
\begin{array}{cccc}
 -\rho & 0 & 0 & 0  \\
 0 & 0 & 0 & 0  \\
 0 & 0 & p & 0  \\
 0 & 0 & 0 & p \\
\end{array}
\right)}.
\end{equation}
One can find this stress tensor using the Israel junction conditions \cite{Israel:1966rt}:
\begin{equation}
T^\mu\,_{\nu}= \frac{M_p^2}{8\pi}\left(\Upsilon^\mu\,_\nu - P^\mu\,_\nu\Upsilon\right)
\end{equation}
where $P$ is the projection operator transverse to the $r$ coordinate, and $\Upsilon_{\mu\nu} = K^-_{\mu\nu} - K^+_{\mu\nu}$.  The extrinsic curvature $K^{\pm}_{\mu\nu}$ of the outer and inner metric are defined as
\begin{equation}
K^{\pm}_{\mu\nu} = P^\alpha\,_\mu P^\beta\,_\nu \nabla_{\alpha} n^{\pm}_{\beta}
\end{equation}
where $n^\pm$ are the unit normal vectors on the outer and inner surfaces.  Plugging in the inner and outer metrics, one can derive the components of $T^\mu\,_\nu$:
\begin{equation}
    \rho = \frac{M_{pl}^2}{4 \pi r_0}\left(\frac{\gamma_{o} - \gamma_{i}}{\gamma_{o} \gamma_{i}} \right)
\end{equation}
\begin{equation}
    p = \frac{M_{pl}^2}{8 \pi r_0} \left( \frac{1}{\gamma_{i}} - \frac{1}{\gamma_{o}} +  \frac{1}{2 r_0^2} \left(\gamma_o h_o - \gamma_{i} h_i\right) \right)
\end{equation}
where the function $h_{i/o} = \left(4 r_0^2 + 2 r_{1i/o} r_{2i/o} - 3 r_0 \left(r_{1i/o} + r_{2i/o} \right) \right)$. With $r_0 = r_{1o}\left(1 - \epsilon \right), \epsilon > 0, r_{1i} > r_{1o}, \text{ and } r_{2i} > r_{2o}$, we have $C > 0$ and $\rho > 0$. Our goal is to create a singular (Planck-density) surface at $r_{1o}$. Accordingly, we work in the limit $\epsilon \rightarrow 0$ where the density approaches: 
\begin{equation}
    \rho \rightarrow \frac{M_{pl}^2}{4 \pi r_{1o}^2} \sqrt{\left(r_{1i} - r_{1o}\right) \left(r_{2i} - r_{1o}\right)}
\end{equation}
and the pressure is divergent and of the form: 
\begin{equation}
    p = \frac{M_{pl}^2}{16 \pi r_{1o}} \left(- \sqrt{\frac{r_{2o} - r_{1o}}{\epsilon r_{1o}}}  + \mathrm{finite \, pieces} \right) 
\end{equation}
Note, the pressure required for the shell is negative. To place the shell at a physical distance $\sim 1/M_{pl}$ from the inner horizon $r_{1o}$ of the exterior black hole, we take $\epsilon \sim {(r_{2o}-r_{1o})}/{(M_{pl}^2 r_{1o}^3)}$, which automatically makes the pressure density of order $M_{pl}^3$.  For the density to also become Planckian, we take $r_{1i}, r_{2i} \sim M_{pl} r_{1o}^2$.   By appropriate choice of coefficients, we can get $|p| < \rho$ so that the shell obeys the dominant energy condition, plausibly enabling the matter on the shell to be provided by   canonical classical fluids. For these parameters, $R_{\mu \nu \lambda\sigma}R^{\mu \nu \lambda \sigma}$ is $\sim M_{pl}^4$ at $r_{1o}$, implying the breakdown of general relativity just within the inner horizon of the exterior black hole. Moreover, the physical length of this region is also $\sim 1/M_{pl}$ {\it i.e.} the location of the singularity, at $r=0$, is a Planck length away from the inner horizon. Note that while we expect the breakdown of GR between $r_{1o} >r > 0$, this solution is a limit of well-defined solutions to Einstein's equations at sub-Planckian densities and curvatures.

The ratio of the charge density (measured in Planck units) to the mass of the shell is $\sqrt{\frac{r_{1i} r_{2i}}{\left(r_{1i} - r_{1o}\right)\left(r_{2i}- r_{1o}\right)} \alpha_{EM}} \gtrapprox 1$. While this ratio is bigger than one, it can be easily accommodated in typical matter - for example, this ratio is $\sim 10^{19}$ for the proton. Another interesting question to ask is if the shell could exist by itself, without the associated inner singular structure. In particular, we can ask if a shell with the calculated charge, density and surface area could exist in free space, wherein its interior is Minkowski space and the exterior is a charged Reissner Nordstrom solution. It can be shown that there are no such static solutions when the mass density of the shell is larger than $M_{pl}^2/r_{1o}$, a density that is much lower than the required Planck density. This shows that this static structure can exist only under the influence of the binding forces exerted by the inner singularity.

The proposed setup above thus allows one to avoid the  troubles caused by the inner horizons of Reissner-Nordstrom black holes. Importantly, the breakdown of general relativity at the scale of the inner horizon is not caused by external perturbations - rather, it is simply caused by charge separation within the singularity itself, leading to an expanded inner singularity that nevertheless matches to the same ADM parameters of the black hole.

\section{Schwarzschild Black Hole}
\label{sec:Schwarzschild}

\begin{figure}[h!]
\includegraphics[width = \textwidth]{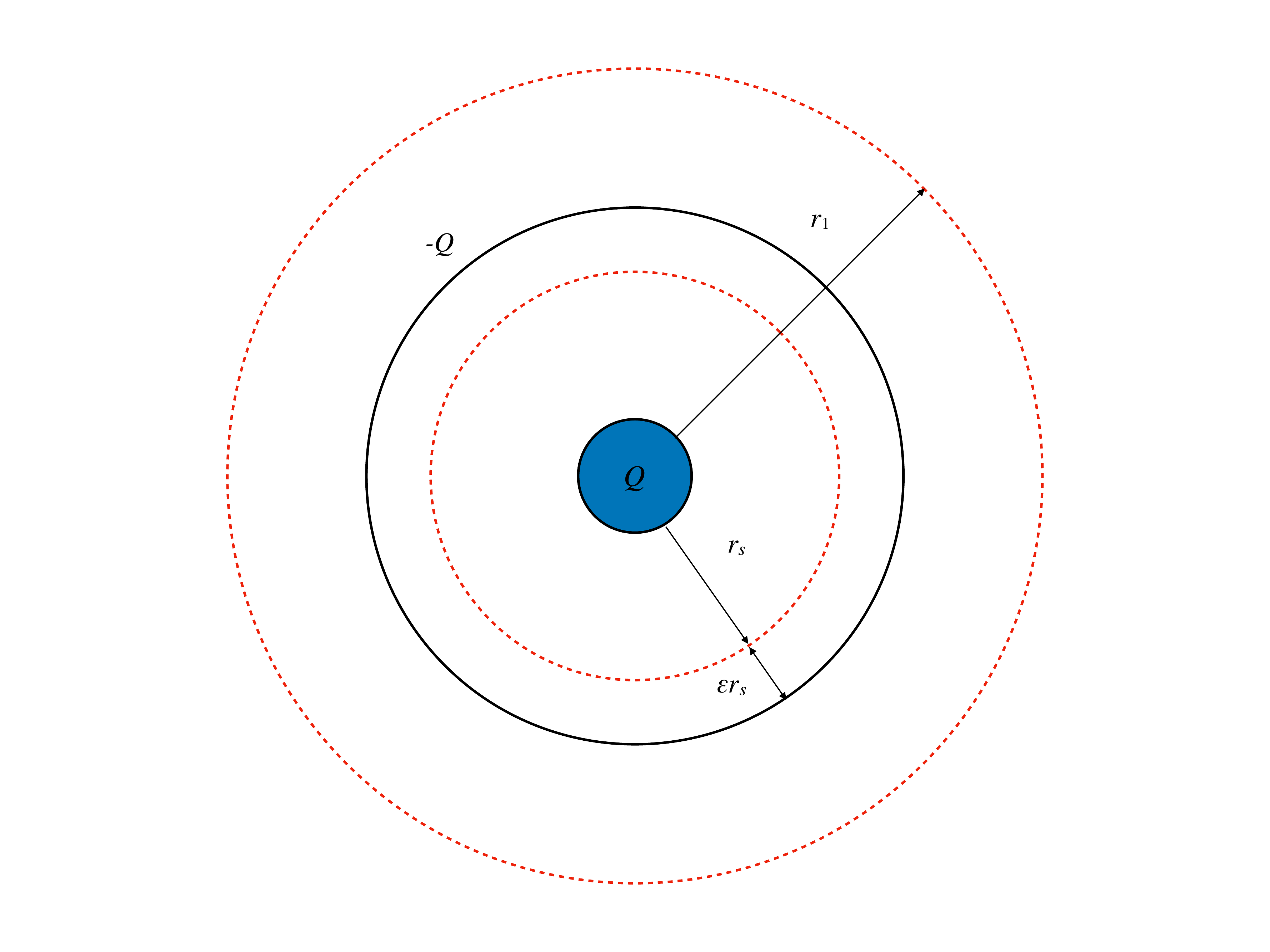}
\caption{The interior of the Schwarzschild black hole is replaced with that of a Reissner Nordstrom black hole of charge Q. The inner horizon $r_1$ of the Reissner Nordstrom black hole is taken to be much larger than the Schwarzschild radius $r_s$ of the target black hole. A shell of charge -Q is placed just outside the Schwarzschild radius $r_s$. The metric outside this shell is the Schwarzschild metric for a black hole of radius $r_s$ while the interior is that of the assumed Reissner Nordstrom black hole. }
\label{fig:BHInterior}
\end{figure}

Now we will take the more radical step and look for solutions with a Planck-density shell just outside a Schwarzschild horizon and look for limits of interior solutions in GR that match on to this boundary.  We then posit how such a naked singularity could be the endpoint of the evolution of a traditional black hole.  Note that this solution can easily be extended to a Reissner-Nordstrom exterior (charged black hole) with a shell placed just outside its outer horizon.

Our setup is described in Figure \ref{fig:BHInterior}. Proceeding as above, we wish to place a firewall just outside the Schwarzschild radius $r_s$ of a black hole. To do so, we assume that the interior of the black hole is described by a Reissner-Nordstrom metric whose inner and outer horizons, $r_1$ and $r_2$, are much larger than the Schwarzschild radius $r_s$ of the target black hole. The metric outside the shell is taken to be the Schwarzschild metric of the target black hole. This of course requires that the charge on the shell be exactly the opposite of the charge on the interior Reissner-Nordstrom black hole. With $r_1 \gg r_s$, this geometry describes a space-time without any horizons. We can thus choose global coordinates $\left(t, r, \theta, \phi\right)$ to cover the entire space-time. The shell that is located at the point $r_0 = r_s \left(1 + \epsilon\right)$ is a time-like surface, with a normal vector pointing radially outward. 
%

We again wish to match these two space-times across the shell using the Israel junction conditions. The metric in the interior $g_{in}$ is taken to be: 
\begin{equation}
\label{eq:SchIn}
g_{in} = - C^2 f_{i}\left(r\right)dt^2 + \frac{dr^2}{f_{i}\left(r\right)} + r^2 d\theta^2 + r^2 \sin^2\theta d\phi^2
\end{equation}
where $f_{i}\left(r\right) =\frac{ \left(r - r_1\right) \left(r - r_2\right)}{r^2}$, with $r_1$ and $r_2$ being the inner and outer horizons of the interior Reissner-Nordstrom black hole and $C$ is again a constant re-definition of the local clocks in the interior in order to facilitate the matching of the metrics across the surface. The exterior metric $g_{out}$ is taken to be the usual Schwarzschild metric: 
\begin{equation}
g_{out} = - f_{o}\left(r\right)dt^2 + \frac{dr^2}{f_{o}\left(r\right)} + r^2 d\theta^2 + r^2 \sin^2\theta d\phi^2 
\end{equation}
where $f_{o}\left(r\right) = \left(\frac{r - r_s}{r} \right)$. Matching these two metrics at the point $r = r_0$ yields
\begin{equation}
    C = \frac{\gamma_{i}}{\gamma_{o}} \quad \mathrm{ where } \quad \gamma_i =\sqrt{ \frac{r_0^2}{\left(r_1 - r_0\right)\left(r_2 - r_0\right)}}, \gamma_o = \sqrt{\frac{r_0}{r_0 - r_s}}
\end{equation} 
%
%
%
%
%
%
The density $\rho$ and pressure $p$ on the shell are
\begin{equation}
\rho = \frac{ M_{pl}^2}{4 \pi r_0} \frac{\left(\gamma_o - \gamma_i \right)}{\gamma_o \gamma_i}
\end{equation}
\begin{equation}
\label{eq:SchP}
p = \frac{ M_{pl}^2}{16 \pi r_0^2} \left( -\gamma_o r_s + 2 r_0\left(\gamma_o - \gamma_i\right) + \gamma_i\left(r_1 + r_2\right) \right)
\end{equation}
Here, the pressure required for the shell can be positive -- this is the only notable difference from the parameters of the shell required for the Reissner-Nordstrom solution above.  With $r_0 = r_s \left(1 + \epsilon\right), \epsilon > 0$, and $r_2>r_1 > r_0$, we have $C > 0$ and $\rho > 0$. Thus, the two space-times can be successfully joined at the shell with positive density matter. We would also like this matter source to obey the dominant energy condition so that it can plausibly be provided by canonical classical fluids. We study this in the limit of small $\epsilon$, so that we can place the shell close to $r_s$. As $\epsilon \rightarrow 0$,
 \begin{equation}
 \rho \rightarrow \frac{M_{pl}^2}{4 \pi r_s^2} \sqrt{\left(r_1 - r_s\right) \left(r_2 - r_s\right)}
 \label{eqndensity}
 \end{equation}
$p$ is divergent and is of the form
\begin{equation}
    p \rightarrow \frac{M_{pl}^2}{16 \pi r_s^2}\left( \frac{r_s}{\sqrt{\epsilon}} + \left(r_1 + r_2 - 2 r_s \right) \sqrt{\frac{r_s^2}{(r_1-r_s)(r_2 - r_s)}} \right)
    \label{eqpressure}
\end{equation}
In the limit $r_1 \sim r_2 \gg r_s$, the finite part of the pressure in \eqref{eqpressure} is clearly sub-dominant and is smaller than the density \eqref{eqndensity}. 
%
%
%

To obey the dominant energy condition (assuming the matter is a perfect fluid along the surface), we need $|p| \lessapprox \rho$. This implies 
\begin{equation}
\epsilon \gtrapprox \frac{r_s^2}{ \left(r_2 - r_s\right) \left(r_1 - r_s\right)}
\label{eqnposition}
\end{equation}
Physically, an observer in the exterior would interpret this shell as being at a physical distance $\gtrsim r_s^2/\sqrt{r_1 r_2}$ outside the Schwarzschild radius $r_s$.

These conditions are clearly easy to satisfy and yield a shell whose stress-tensor satisfies the dominant energy condition that allows us to match the interior Reissner-Nordstrom metric with the exterior Schwarzschild metric. This is a horizon-free geometry and thus provides an example of a new kind of naked singularity. The charge on the central Reissner-Nordstrom singularity is $Q = M_{pl} \sqrt{\frac{r_1 r_2}{\alpha_{EM}}}$. Thus, in Planck units, the charge-to-mass ratio on the shell is $\sim \frac{Q}{4\pi r_s^2 \rho}\sim\sqrt{\frac{r_1 r_2}{\left(r_1 - r_s\right)\left(r_2 - r_s\right)}} \frac{1}{\sqrt{\alpha_{EM}}} \gtrapprox 1$, similar to the previous example. 

There are a variety of choices of the parameters $\epsilon, r_1 \text{ and } r_2$ that yield stress tensors which satisfy the dominant energy condition with a reasonable ratio of the charge to mass density on the shell. For example, when $r_2 \gtrapprox r_1 \gtrapprox r_s$ ({\it i.e.}, all the same order of magnitude), $\rho\sim p \sim M_{pl}^2/r_s$.  We are however interested in higher (Planckian) density matter as we discuss below.  This is achieved by setting $\sqrt{r_2 r_1}\sim r_s^2 M_{pl}$ (as long as the shell is inside $r_1$).  Remarkably, this condition on $r_2 r_1$ also sets the physical distance from the shell to the inner singularity to be a Planck length!  Thus, this shell is Planck density, is macroscopic in size, and is yet a Planck length away from the center of the black hole.  In effect, the entire black hole lives at its surface!

Finally, $p\sim\rho$ when either $r_2\gg r_1 \gtrapprox r_s$ or the shell is a Planck length outside of $r_s$ ({\it i.e.}. $\epsilon\sim 1/(r_s M_{pl})^2)$).  We will focus on the latter when $r_2\sim r_1$, as the part of parameter space with small $r_1$ can also lead to negative-mass Schwarzschild solutions, which we will discuss in Section \ref{sec:NM}.  For completeness, we note that there are also regions with $p\ll\rho$ when $\epsilon\sim {\cal O}(1)$ and $r_2\gtrapprox r_1\gg r_s$.


Our solution yields a naked singularity matching on to the Schwarzschild exterior. The Planckian density shell would naively imply a large mass for the black-hole. But, because of the nature of the interior Reissner-Nordstrom solution, this does not happen - the ADM mass of the black hole is that of the exterior Schwarzschild solution. However, we have not provided a plausible path towards forming such a naked singularity. In particular, given a Schwarzschild black hole, how could it evolve into the structure depicted in figure \ref{fig:BHInterior}? 

\subsection{Formation}

Consider a canonical Schwarzschild black hole that was formed due to the collapse of matter. Naively, inside the Schwarzschild horizon of this black hole, the singularity becomes a point in time.  This description is valid only when general relativity holds. It is easily verified that in the Schwarzschild geometry, the scalar $R_{\mu \nu \lambda \sigma}R^{\mu \nu \lambda \sigma}$ becomes $\sim M_{pl}^4$ at  $r \sim M^{\frac{1}{3}}/M_{pl}^{\frac{4}{3}}$ from the singularity (a Planck time from the singularity at $r=0$). The geometric description of space-time is thus incorrect for $r \lessapprox M^{\frac{1}{3}}/M_{pl}^{\frac{4}{3}}$. The singularity should be viewed as being spread over this region. We would like our charged shell to emerge from this singular region and make its way to the horizon. 

If the shell was made of matter with sub-Planckian densities, the motion of the shell must be consistent with general relativity. In this case, the shell cannot propagate from the singular region to the Schwarzschild horizon without violating causality. The only way such a motion could be possible is if the shell itself was made out of Planckian density.
In this case, the local motion of the elements of the shell need not obey general relativity since this motion can be influenced by local (Lorentz invariant but equivalence principle violating) higher dimension operators on the shell. While these operators cannot affect the center of mass motion of the shell, they can indeed affect its expansion.  This is our motivation to choose parameters so that the shell itself is Planckian. 


The singular nature of the shell and the interior geometry is a key difference between our picture and other proposals such as fuzzballs \cite{Mathur:2009hf} where the mass of the black hole is pushed outside the event horizon in the form of low density matter. In this case, the collapsing matter would have to transition into the fuzzball state before the formation of the event horizon and it is unclear how this process could occur. In our picture, the collapsing matter can hit a singular region and then subsequently re-emerge as a singular shell that is nevertheless still able to get to the horizon. 

With these parameters described above, the structure described in figure \ref{fig:BHInterior} is a shell at Planckian density (a firewall) at a distance $\sim 1/M_{pl}$ outside the Schwarzschild radius of the exterior black hole.  The geometry describes a naked singularity. The physical radius of this shell is also $\sim 1/M_{pl}$ and the singularity is thus at the Schwarzschild horizon.  The complete breakdown of general relativity within the shell implies that the evolution of the shell from the initial Schwarzschild singularity to the firewall state could happen without upsetting the asymptotic ADM mass of the black hole. We note that this setup also provides a natural reason for the scaling of the black hole entropy with its area \cite{PhysRevD.7.2333}: in this case, the length of the interior is $\sim 1/M_{pl}$ while the area of the shell is still equal to the usual surface area of a Schwarzschild black hole. 

An unsatisfactory aspect of our construction is that the shell must be placed outside the Schwarzschild horizon. We cannot place such a shell inside the Schwarzschild horizon and apply the Israel junction conditions to obtain the stress tensor on the shell since surfaces of constant $r$ are space-like in this region. Thus, this matching cannot be performed within General Relativity. If the shell and the region enclosed by it are singular, such a matching might be possible in a quantum theory of gravity. Solutions with shells in the interior could lead to a complete description of the formation of the firewall. But, without such a theory, we cannot show that the singular mass could develop in the way we would like without changing the ADM parameters of the theory. However, the fact that such solutions exist within General Relativity when the shell is outside the horizon makes us optimistic that quantum gravity will exhibit similar behavior.

\section{Negative Mass Schwarzschild}
\label{sec:NM}

In the previous section, we found solutions to Einstein's equations which look like exterior Schwarzschild metrics that end with a Planck-density shell at the horizon and a central singularity a Planck-length away.  One can ask if it is possible to use the same tools to construct  negative-mass Schwarzschild solutions.  Indeed, by simply taking $r_s\rightarrow -r_s$ in Equations \eqref{eq:SchIn} - \eqref{eq:SchP}, one will discover viable solutions with $\rho\sim p\sim M_{pl}^3$.  While the formation of such solutions from asymptotic normal matter is impossible, they potentially pose a problem of quantum instability of the vacuum.

Naively, one worries about catastrophic decay of the vacuum into pairs of positive and negative mass solutions \cite{Cline:2003gs}. If such a process is possible (and it is not clear that it is), it should be wildly suppressed by form factors.  However, any non-zero rate becomes divergent in a Lorentz-invariant background as the Lorentz group is non-compact and the phase space is infinite. Thus, the rate must be regulated with a Lorentz-violating cutoff. Of course our universe is not Lorentz-invariant, and if, for example, the universe has a finite lifetime or has a finite volume, it could be enough to suppress the decay to acceptable levels.

Another possibility is that the negative-mass solutions are indicative of restrictions on the Planck-density matter.  Indeed, there are places in the $\rho-p$ parameter-space with both positive and negative Schwarzschild solutions, but where the space-time curvature in the negative mass case is super-Planckian.  For example, take $r_2\sim 4 r_1$ and $\rho\simeq M_{pl}^3/4\pi\simeq 4 p$.  This can be achieved by taking $r_s \rightarrow r_0 - 1/(M_{pl}^2 r_0)$ for the positive mass black hole and $r_s \rightarrow M_{pl}^2 r_0^3 \gg r_0$ for the negative mass black hole.\footnote{We could take  $r_0 \lesssim 1/M_{pl}$ - but for a macroscopic black hole, this point lies well within the region  $ r \lesssim \left( r_1 r_2\right)^{1/4} / \sqrt{M_{pl}}$ where general relativity breaks down.}  The shell is apparently identical in both cases.  However, for the positive Schwarzschild case, Riemann squared just outside of the shell is $R^{\alpha\beta\gamma\delta} R_{\alpha\beta\gamma\delta} \sim - 12/r_0^4$, whereas for the negative Schwarzschild solution, the same quantity is $- 12 M_{pl}^4$, which is numerically outside the regime of validity.  Thus, the latter may not be a solution in the full theory and if only a certain class of shells are allowed, the negative Schwarzschild examples may not exist.

\section{Discussion and Conclusions}

We have shown that charge and mass separation within the interior of a black hole can lead to dramatic changes to the interior geometry while leading to the same exterior. In particular, the interior geometry can lead to the existence of singular firewalls at the event horizon of a  Schwarzschild (or Reissner-Nordstrom) black hole and the inner-horizon of a Reissner-Nordstrom black hole. Traditionally, efforts to solve the  problems of the inner horizon of a  Reissner-Nordstrom black hole have been detached from attempts to solve the black hole information problem associated with the Schwarzschild event horizon. It is clear that the  problems of the inner horizon of a Reissner-Nordstrom black hole cannot be solved without a change to the classical geometry of the black hole interior - in particular, a singularity should exist at the inner horizon. We have shown how such a singularity might form in a region that naively has a low curvature scale. This singularity does not require external perturbations - it can arise simply from the physics of the interior. This picture also naturally leads to the existence of a similar structure for Schwarzschild black holes, with a singularity at the event horizon. In both these cases, the singular interior has a short physical length ($\sim 1/M_{pl}$) - but the surface area is the same as that of the exterior solutions. This provides a natural physical picture for the scaling of black hole entropy with area - the black hole should simply be thought of as effectively a 2D object, with a large surface area and a small interior volume. 

It is likely that similar structures also exist in Kerr black holes  \footnote{Moreover, there might be other singular solutions \cite{shell} that could conceivably be adapted to our picture of black hole evaporation.}. While we have not attempted to find such solutions, this is a reasonable expectation since Kerr black holes also possess troublesome inner horizons. In fact, it is plausible that the exterior Schwarzschild solution could be produced by the combination of a rapidly rotating inner singularity and a counter rotating outer singular shell. This is in fact the main message of this paper: the interior of a Schwarzschild black hole could be replaced by a more complex geometry, leading to a firewall at the event horizon, without change to the external parameters of the black hole. The fact that Kerr and Reissner-Nordstrom geometries need to have a complex geometry at their inner horizon makes it plausible that the Schwarzschild black hole also has a similar structure.

An important deficiency of our work is that we have not identified the micro-physics of the shell necessary to match the solutions together. While we have shown that the shell obeys the dominant energy condition, without a microscopic theory, we are unable to determine its stability. Indeed, an unstable shell might even permit a new way for the black hole to decay, for example, through an explosion of an unstable shell rather than slow Hawking evaporation. Moreover, while we have shown that these singular geometries exist, we have also not investigated the dynamics that could lead to their formation in a singular environment. Such a treatment would require a theory of quantum gravity, and it would be interesting to see if frameworks such as string theory could support the formation and evolution of such structures. 

The existence of a firewall in the inner horizon of a Reissner-Nordstrom black hole cannot be observed by an observer in the exterior. However, if there is a firewall outside the horizon of a Schwarzschild black hole, there are likely to be observational signatures. First, since these are geometries without a horizon, the no hair theorems will be violated, potentially leading to observational signatures in experiments such as LIGO and the Event Horizon telescope. Second, collisions between such black holes could lead to the production of electromagnetic radiation - with the advent of multi-messenger astronomy, follow up observations of black hole mergers might reveal novel signatures that are not expected within General Relativity. There have also been speculations that the existence of firewalls could lead to echoes of the gravitational waves produced in the mergers of black holes \cite{Abedi:2016hgu} - our metric provides a concrete framework to quantitatively estimate these effects. A major concern for this observational program is that we do not know when the interior of the black hole would evolve into the firewall. Logically, such an evolution is necessary only within the Page time, a time scale that is much longer than the age of the universe for a solar mass black hole. However, there is reason to be optimistic - the firewall needs to form rapidly in the interior of a Reissner-Nordstrom black hole in order to avoid the problems of the inner horizon of such a black hole. If both of these problems are to be solved by the same mechanism, it is reasonable to hope that the formation time for a firewall in the Schwarzschild case is also similarly short.

Our picture also has significant import for particle physics. In our scenario, it is reasonable that the destruction of the black hole happens due to the physics of the firewall rather than Hawking evaporation. While Hawking evaporation is an instability of black holes and shows that they cannot be eternal, it is possible that firewalls lead to parametrically faster decay of the black hole. Such a decay could conceivably conserve particle properties such as global charges that may have been associated with the matter whose collapse produced the black hole. This is further bolstered by the fact that our solutions are naked singularities without associated no hair theorems. A fundamental feature of our solution is that the binding energies in the interior of the black hole are able to dramatically cancel interior energies leading to a parametrically smaller exterior mass for the black hole. This occurs without any obvious symmetry. It is interesting to ask if such strutures are also possible in gauge theories. Gauge theories do contain examples of massless states that are nevertheless composed of high energy degrees of freedom. For example, in confining theories such as QCD and the composite axion, spontaneous symmetry breaking can lead to light goldstone bosons where the high energies of the particles within the composite state is cancelled to high accuracy. While symmetry is a natural way to enforce this cancellation, our black hole solutions suggest that there might be other dynamical structures that could also lead to dramatic cancellations between interior energies and binding energies, leading to light states without any obvious symmetry. If such theories are found, they could potentially lead to new paths to solve the hierarchy problem. 

\section*{Acknowledgements}
We thank Nima Arkani-Hamed, Ibrahima Bah, Ted Jacobson, Shamit Kachru, Nemanja Kaloper, Jared Kaplan and Raman Sundrum for useful conversations. DK was supported in part by the NSF under grant PHY-1818899.  SR was supported in part by the NSF under grants PHY-1638509 and PHY-1507160, the Alfred P. Sloan Foundation grant FG-2016-6193, the Simons Foundation Award 378243 and the Heising-Simons Foundation grant 2015-038. SR also acknowledges the support of the Bearden Professorship at Johns Hopkins University when this work was initiated. 

\bibliography{references}
\end{document}